\documentclass[prb,twocolumn,showpacs]{revtex4}
\usepackage{graphicx}
\usepackage{amsmath}
\usepackage{amssymb}
\usepackage{gensymb}
\usepackage[sort&compress]{natbib}

\begin{document}
\author{C.R.~Dean$^{1,2,\dag}$, A.F.~Young$^{3,\dag}$, P.~Cadden-Zimansky$^{3,4}$, L.~Wang$^{2}$, H.~Ren$^{3}$, K.~Watanabe$^{5}$, T.~Taniguchi$^{5}$, P.~Kim$^{3}$, J.~Hone$^{2}$, K.L.~Shepard$^{1}$}
\affiliation{$^{1}$Department of Electrical Engineering, Columbia University, New York, NY, 10027, USA}
\affiliation{$^{2}$Department of Mechanical Engineering, Columbia University, New York, NY, 10027, USA}
\affiliation{$^{3}$Department of Physics, Columbia University, New York, NY, 10027, USA}
\affiliation{$^{4}$National High Magnetic Field Laboratory, Tallahassee, Florida, 32310, USA}
\affiliation{$^{5}$Advanced Materials Laboratory, National Institute for Materials Science, 1-1 Namiki, Tsukuba, 305-0044, Japan}
\title{Multicomponent fractional quantum {H}all effect in graphene}

\begin{abstract}We report observation of the fractional quantum Hall effect (FQHE) in high mobility multi-terminal graphene devices, fabricated on a single crystal boron nitride substrate.  We observe an unexpected hierarchy in the emergent FQHE states that may be explained by strongly interacting composite Fermions with full SU(4) symmetric underlying degrees of freedom. The FQHE gaps are measured from temperature dependent transport to be up 10 times larger than in any other semiconductor system. The remarkable strength and unusual hierarcy of the FQHE described here provides a unique opportunity to probe correlated behavior in the presence of expanded quantum degrees of freedom.
\end{abstract}


\maketitle

The fractional quantum Hall effect\cite{Tsui1982, Laughlin1983, Halperin1983, Jain1989} (FQHE) was first described by a single component model with no internal degrees of freedom\cite{Laughlin1983}.  Multicomponent systems, however, necessitate a generalization of the theory to allow for ordering in the space of the additional degeneracy. In graphene, the structure of the honeycomb lattice endows the wavefunctions with an additional quantum number, termed valley isospin, which, combined with the usual electron spin,  yields four-fold degenerate, SU(4) symmetric Landau levels\cite{Novoselov2005, Zhang2005}. This additional symmetry fundamentally modifies the FQHE and, in graphene, is conjectured to produce new incompressible ground states that have no analogue in conventional systems~\cite{Toke2006,Yang2006,Goerbig2007,Nomura2006,Apalkov2006,Khveshchenko2007,Toke2007,Shibata2008,Shibata2009, Papic2010}.

\begin{figure}[b]
	\begin{center}
	\includegraphics[width=1\linewidth,angle=0,clip]{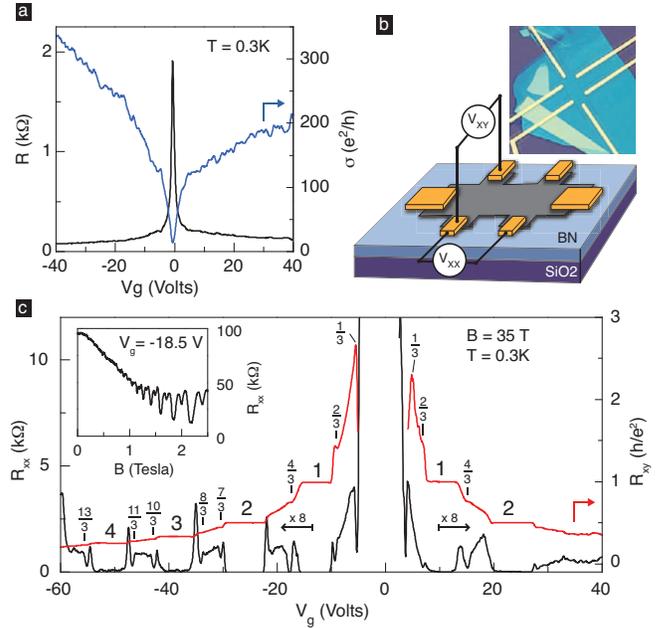}
			\caption{\textbf{Magnetotransport.} (a) Resistance and conductivity measured at zero magnetic field. (b) Schematic of the etched hall bar device. Inset: Optical image.  (c)Magnetoresistance (left axis) and Hall resistance (right axis) versus gate voltage acquired at $B=35$~T. Inset shows SdH oscillations at $V_{g}=-18.5$~V.}
		\label{fig1}
	\end{center}
\end{figure}

\begin{figure*}[th] 	
	\begin{center} 	
		\includegraphics[width=0.90\linewidth,angle=0,clip]{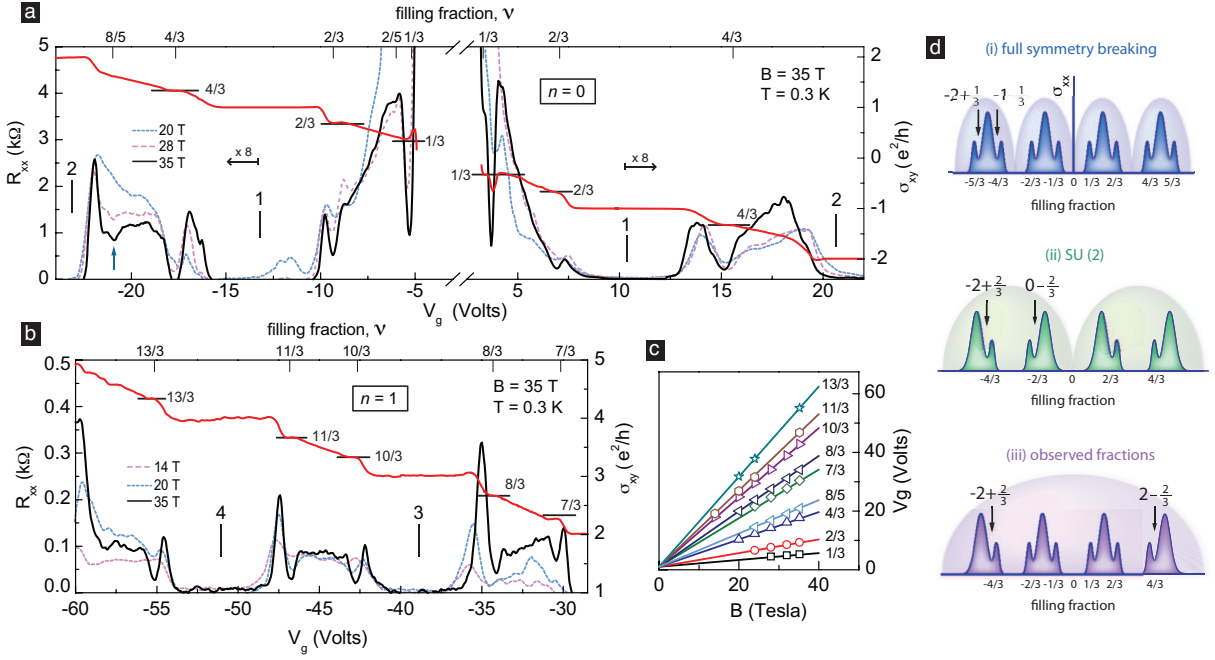}
		\caption{\textbf{Fractional quantum Hall effect.} (a),(b) Magnetoresistance (left axis) and Hall conductivity (right axis) in the $n=0$ and $n=1$ Landau levels at $B=35$~T. $T\sim$0.3~K. (c) Fan diagram showing  the resistance minima at different magnetic fields for the FQHE states labeled in (a) and (b). Lines correspond to the positions calculated from the relation  $B_{\nu}=\frac{1}{\nu}n_{e}h/e$, where $n_{e}=C_{g}(V\!-\!V_{o})/e$ is the carrier density and $C_{g}=1.09\times10^{-4}$~F/m$^{2}$ the capacitive coupling to the doped Si back gate, determined independently from low $B$ Hall measurements.
(d) Cartoon depiction of the FQHE hierarchy observed in our sample. Shown is the expected trend for graphene assuming (i) explicit breaking of all internal degeneracies due to, for example, coupling to external fields, (ii) explicit breaking of one degeneracy with a remaining SU(2) symmetry and (iii) schematic diagram of experimental observation. Arrows label example particle-hole conjugate pairs in each scenario.}
		\label{fig2}	
	\end{center}
\end{figure*}
 
The anomalous quantum Hall effect in graphene~\cite{Novoselov2005,Zhang2005}, in which Hall plateaus appear at $\nu=\pm4(n+\frac{1}{2})$ for Landau level index $n=0,1,2\ldots$, can be understood from a theory of noninteracting, massless Dirac fermions in which the $\frac{1}{2}$ offset arises from the linear dispersion and the four-fold level degeneracy reflects the two spin and two valley degrees of freedom.  Each LL is SU(4) symmetric, a fact not altered by the inclusion of long-range Coulomb interactions between quasiparticles in graphene. Symmetry breaking effects, such as Zeeman coupling and contact interactions, are exceedingly weak at experimentally realizable magnetic fields. Even at $B=35$~T, the ratio between the Zeeman and Coulomb energies is only $E_{Z}/E_{C}\sim0.01\epsilon$, ($\epsilon$ is the dielectric constant), and lattice scale interactions are only $a/l_B\sim0.06$ ($a$ is the lattice constant and $l_{B}=\sqrt{h/eB}$ is the magnetic length).  Neglecting these terms, exchange interactions can favor ground states with a spontaneous SU(4) ferromagnetic polarization~\cite{Nomura2006,Yang2006}, leading to the observation~\cite{Zhang2006,Du2009,Bolotin2009, Dean2010} of the IQHE at all integer $\nu$ in the cleanest samples.  At fractional filling, the expanded symmetry of the LLs opens up the possibility of observing novel many-body states that are distinct from conventional one and two component systems~\cite{Toke2007,Goerbig2007}.

In this Letter we report on measurements of the FQHE in a single-layer graphene sample fabricated on a hexagonal boron nitride substrate (see Methods).   Fig.~\ref{fig1}a shows the zero-field resistance and corresponding conductivity, acquired at T$\sim$300~mK.   The Hall mobility at high density is $\sim$30,000~cm$^{2}$/Vs; the charged-impurity mobility, which dominates at low density, is in excess of 100,000~cm$^{2}$/Vs as determined by fitting straight lines to the linear portion of the conductivity; the charge inhomogeneity, estimated from the resistivity peak width at the charge neutrality point (CNP), is of order 10$^{10}$~cm$^{-2}$.  All three metrics indicate this sample to be of exceptionally high quality and consistent with previous measurements of similar graphene/h-BN devices at low magnetic fields~\cite{Dean2010}.

Magnetoresistance ($R_{xx}$) and corresponding Hall resistance ($R_{xy}$) acquired by varying the gate voltage at a constant B field of 35~T are shown in Fig.~\ref{fig1}c. A complete lifting of the four-fold degeneracy is seen in both the $n=0$ and $n=1$ LLs, with quantized Hall plateaus and magnetoresistance zeroes appearing at all accessible integer fillings.  Signatures of symmetry-breaking at half-filled LLs appear in the Shubnikov-de Haas (SdH) oscillations at fields as low as 1~T (inset of Fig. \ref{fig1}c), with full breaking of the four-fold symmetry observed at fields less than 5~Tesla.  The most remarkable feature of this sample is the emergence of the FQHE throughout (labeled in Fig.~\ref{fig1}c).  In the remainder of this Letter, we focus our attention on the FQHE states with analysis of the IQHE states to be given elsewhere.

In Fig.~\ref{fig2}a,b detailed plots of the $n=0$ and $n=1$ LLs are given. For clarity, the Hall conductivity, calculated from the tensor relation $\sigma_{xy}=R_{xy}/(R_{xy}^{2}+(w/l)R_{xx}^2)$, is shown where $w/l$ is the aspect ratio of the Hall bar.  We unambiguously observe hallmark features of the FQHE, namely quantization of $R_{xy}$ to values of $\frac{1}{\nu}h/e^{2}$ concomitant with minima in $R_{xx}$, at fractional filling factors $\nu=\frac{1}{3},\frac{2}{3}$ and $\frac{4}{3}$ in the $n=0$ LL and at $\nu=\frac{7}{3},\frac{8}{3},\frac{10}{3},\frac{11}{3}$ and $\frac{13}{3}$ in the $n=1$ LL. Additionally, a weak minimum in $R_{xx}$ at $\nu=\frac{8}{5}$ is suggestive of an emerging FQHE state at this filling. With the exception of  $\frac{1}{3}$, $\frac{2}{3}$ and $\frac{7}{3}$, all observed Hall plateaus are within $1\%$ of their expected value.
At $\frac{1}{3}$ and $\frac{2}{3}$, the Hall plateaus are ``N'' shaped and do not exhibit exact quantization, possibly due to their proximity to the insulating $\nu=0$ state~\cite{Checkelsky2008}.

The  FQHE hierarchy observed in the $n=0$ LL (Fig.\ref{fig2}a) can be interpreted in the context of a composite fermion (CF) theory with underlying SU(4) symmetry. In the CF picture, the FQHE is understood by mapping the system of strongly interacting electrons in a large applied $B$ field to a system of weakly interacting CFs consisting of an electron bound to $2p$ magnetic flux vortices~\cite{Jain1989}.  The CFs move in a reduced effective magnetic field $B^{*}=B-B_{1/2p}$, so that the FQHE at $\nu=\frac{m}{2pm\pm1}$ can be viewed, for given values of $p$, as a CF IQHE where $m$ labels the number of filled CF Landau levels (termed as `Lambda' levels). In conventional systems, such as
GaAs, Coulomb interactions heavily mix spin branches within a LL, and the resulting doublet respects an approximate SU(2)
symmetry\cite{Halperin1983} such that even-numerator states (\textit{e.g.} $\nu=\frac{2}{3}$ and $\frac{2}{5}$)  consist of SU(2) spin-singlet states of fully filled, two-fold degenerate, CF LLs.  Odd numerator states, in contrast, correspond to spontaneously spin-polarized quantum Hall ferromagnetism arising from residual interactions between the CF quasiparticles~\cite{Yang2006}. In generalizing this CF hierarchy to graphene, three possible scenarios are conceivable according to the degree of symmetry preserved in the underlying LL of real electrons (see fig. \ref{fig2} d): (i) all degeneracies are explicitly broken, for example, by coupling to external fields; (ii) only one of spin or valley isospin degeneracy is broken, preserving an SU(2) symmetry in the remaining degenerate space; or (iii) the full degeneracy is preserved, leading to an emergent SU(4) symmetry in the combined spin-isospin space.  

The scenarios can be distinguished by particle-hole symmetry.  In each scenario, we expect particle-hole conjugate FQHE states within a single LL to have similar spin textures and gaps. For example, in scenario (i) the $\frac{1}{3}$,$\frac{2}{3}$ and $\frac{4}{3}$,$\frac{5}{3}$ are conjugate pairs and so should show similar energy gaps, whereas, for the SU(2) case (scenario (ii)) the $\frac{5}{3}$,$\frac{1}{3}$ and $\frac{4}{3}$,$\frac{2}{3}$ states represent conjugate pairs in the doubly degenerate levels. For SU(4) (scenario (iii)) only a symmetry over the full LL is expected. The appearance of $\pm\frac{1}{3}$, $\pm\frac{2}{3}$ and $\pm\frac{4}{3}$, and the absence of $\pm\frac{5}{3}$, is inconsistent with both scenario (i) and (ii) which require simultaneous observation of either $\pm\frac{4}{3}$ and $\pm\frac{5}{3}$ or $\pm\frac{1}{3}$ and $\pm\frac{5}{3}$, respectively~\cite{Dunford95,Lai04}.  This suggests SU(4) multicomponent FQHE as the scenario that best fits our experimental observations where clear symmetry exists only across the full $n=0$ LL.

In an SU(4) symmetric LL singlet states are expected at fractional fillings with numerator $m=4$, with residual CF interactions responsible for the emergence of the FQHE at all other fillings in the same way that interactions between real electrons drive the IQHE at $\nu\neq\pm4(N+\frac{1}{2})$. The appearance of the $\frac{4}{3}$ and (possibly) $\frac{8}{5}$ in the absence of $\pm2\mp\frac{4}{9}$ indicates that the spontaneously broken CF states are more robust than SU(4) spin singlets.  This suggests that CFs are not very good quasiparticles in graphene, in that they continue to interact strongly.  It is important to note that in the CF construction, the filling factors are measured relative to the empty LL so that in graphene, due to the Berry phase shift, the analog of $\frac{1}{3}$ filling in a traditional 2DEG is found at $-2+ \frac{1}{3} =-\frac{5}{3}$ for electrons (or at $2-\frac{1}{3}=\frac{5}{3}$ for holes).  The $\frac{1}{3}=2-\frac{5}{3}$ is thus \textit{not} the graphene analog of the Laughlin state at $\nu=\frac{1}{3}$ in GaAs, and may possess a different spin texture~\cite{Papic2010}.  The reasons for its strength relative to the absent $\frac{5}{3}=2-\frac{1}{3}$ are not clear, but may be associated with this different ordering in the SU(4) space. 

 \begin{figure}[t] 	
 	\begin{center} 	
 		\includegraphics[width=1\linewidth,angle=0,clip]{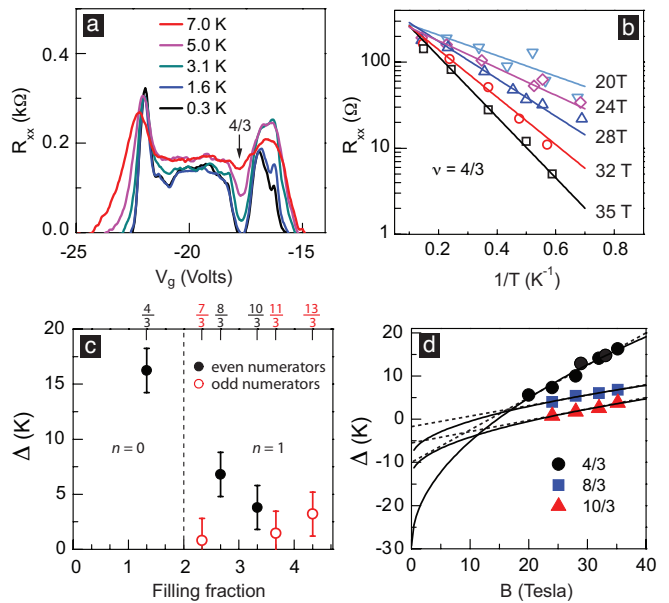}
		\caption{\textbf{Energy Gaps.} (a) Magnetoresistance versus temperature near $\nu=\frac{4}{3}$ at $B=35$~T. (b) Ahhrenius plot of the $R_{xx}$ minimum at $\nu=\frac{4}{3}$ for several magnetic fields. Energy gaps versus  filling factor at $B=35$~T and versus perpendicular magnetic field are shown in (c) and (d). Dashed and solid lines in (d) are fits assuming linear and $\sqrt{B}$ dependence, respectively. }
		\label{fig3} 	
	\end{center}
\end{figure}

In contrast, the second LL (SLL) shows nearly all multiples of $\frac{1}{3}$ within the experimental range, suggesting a different CF interaction strength. For example, the $\frac{7}{3}=2+\frac{1}{3}$ state, the SLL analogue of the $\frac{5}{3}$, is present.  The $\frac{11}{3}=2+\frac{5}{3}$ and $\frac{10}{3}=2+\frac{4}{3}$, analogues of the $\frac{1}{3}$ and $\frac{2}{3}$ are also both well developed.  Observation of the FQHE at $\nu=\frac{13}{3}$ represents, to our knowledge, the first unambiguous report of the FQHE at a filling factor $\nu>4$.  This may be attributed in part to the fact that the FQHE states between $4<\nu<6$ continue to belong to the $n=1$ LL due to the four-fold symmetry of single-particle LLs in graphene.We also note that there is no evidence of any even denominator FQHE, despite the fact that the energy gap at $\nu=\frac{5}{2}$ is typically comparable to the $2+\frac{m}{3}$ states in GaAs~\cite{Toke2008}.

The gap energies of the FQHE states were measured from the temperature dependence of the $R_{xx}$ minima in the thermally activated regime $R_{xx}\propto e^{\Delta/2k_{B}T}$, where $\Delta$ is the energy gap, $k_{B}$ is Boltzmann's constant and $T$ is the electron temperature (Fig. \ref{fig3}). For all observed FQHE states, the gap values are remarkably large, reaching as high as $\sim16$~K at $\nu=\frac{4}{3}$. This compares to the highest mobility GaAs samples where the strongest $\frac{1}{3}$ state is experimentally unobservable above only a few Kelvin. In the $n=1$ LL, the gaps at $\frac{8}{3}$ and $\frac{10}{3}$ are more than an order of magnitude larger than at the same total filling in GaAs~\cite{Kumar2010}.  The enhanced gaps in our sample result in part from the lower dielectric screening and near-zero width of the 2DEG, both of which increase the effective strength of the electron interactions that give rise to the FQHE.  Interestingly, the even numerator and odd numerator states exhibit opposite trends, with the even-numerator decreasing with increasing LL index while the odd-numerator states actually grow in strength. Although the proximity of the $\frac{1}{3}$ state to the insulating state does not allow us to measure $\Delta_{1/3}$, recent measurements performed on multi-terminal suspended graphene indicate that $\Delta_{1/3}$ is indeed larger than any gap values measured in this device~\cite{Fereshte2010}. The trend in the gap scaling may be a consequence of the relativistic dispersion relation, which modifies the electron interactions in graphene~\cite{Goerbig2006,Apalkov2006}.  The fact that the even numerator gaps are greater than that of the odd numerator is qualitatively consistent with the broken symmetry IQHE of real electrons, where the even integer states emerge from the otherwise four-fold degenerate LLs with stronger gaps than the odd-integer states~\cite{Zhang2006}.

Measurement of the even-numerator gaps over a range of perpendicular fields is shown in Fig.~\ref{fig3}d. In the simplest picture of spinless, non-interacting composite Fermions, the FQHE energy gaps are set by the Coulomb interaction between real electrons and therefore exhibit a $\sqrt{B}$ dependence with magnetic field.  However, residual CF interactions that give rise to spin-textured charged excitations can result in a linear dependence~\cite{Dethlefsen2006}.  A question of fundamental interest is whether incompressible states at fractional filling in graphene support such skyrmionic excitations, and how these relate to the appropriate underlying symmetry.  The solid and dashed lines in Fig.~\ref{fig3}d are attempts to fit the data using a square root and linear dependence, respectively. Both fit the data equally well, making it impossible at present to distinguish between the two dependences.  Using the two y-axis intercepts as upper and lower bounds on disorder, the intrinsic gap of the $\frac{4}{3}$ state, $\Delta_{i}=\Delta_{meas}-\Gamma$ with $\Gamma$ the disorder-induced LL broadening, is determined to be 0.04-0.06~$e^{2}/\epsilon l_{B}$ in Coulomb energy units. Here we have taken the effective dielectric constant for graphene to be $\epsilon\sim5$, calculated from the internal electron screening due to interband transitions\cite{Ando2006}, where the external dielectric environment is taken as an average of vacuum ($\epsilon_{vac}=1$) and the underlying BN substrate ($\epsilon_{BN}\sim3.5$).

In conclusion, we report the first unambiguous observation of the fractional quantum Hall effect in graphene in both the lowest and second Landau levels. The observed fractions are most consistent with the expected SU(4) symmetry of the single-particle level being broken spontaneously at all fillings at which the FQHE is observed, confirming the large strength of electronic interactions in graphene and suggesting the possibility of observing novel spin textures with no analog in other single-layer quantum Hall systems. In the $n=0$ LL, measurement of the intrinsic energy gap at $\frac{4}{3}$ is found to be in good agreement with theoretical calculations.  In the $n=1$ LL the analogue of the $\frac{1}{3}$ and $\frac{2}{3}$ energy gaps in the degenerate LL are measured to be as much as 10 times larger than have been reported for the highest quality GaAs samples.

\bigskip

\section{Methods}
To fabricate the graphene--on--h-BN device a similar mechanical transfer technique to that described in Ref.~\onlinecite{Dean2010} was used but with the water soluble sacrificial layer replaced by a polyvinyl alcohol (PVA) layer.  This allowed mechanical peeling of the PMMA membrane without the need for exposing the graphene/PMMA substrate to a water bath, thereby achieving a fully dry transfer method. Electrical leads consisting of a Cr/Au metal stack were deposited using standard electron-beam lithography after which the sample was etched into an approximately square Hall bar by exposure to oxygen plasma (Fig. \ref{fig1}b). Four-terminal transport measurements were performed using a lock-in amplifier at 17~Hz with a 10~nA source current.  The sample was measured in a 35~T resistive magnet and $^{3}$He cryostat (sample in vapour). We note that there is an apparent asymmetry in the transport, with p-type carriers showing better developed quantum Hall effect features/higher mobility than n-type carriers, particularly in the second LL.  This higher mobility for p-type carriers is not a systematic trend, as we have observed higher mobility for n-type carriers in other samples.  Therefore, while data acquired at negative gate voltages (p-type carriers) is discussed primarily, we expect our results should be consistent with samples where n-type carriers experience comparable disorder.

\bibliography{Refs}

\noindent
$^{\dag}$These authors contributed equally to this work.
\bigskip

\noindent
Reprints and permission information is available online at http://npg.nature.com/reprintsandpermissions/. Correspondence and requests for materials should be addressed to PK.

\bigskip
\section{Acknowledgments}
We thank J.~K. Jain, C. Toke, M.~O. Goerbig and M. Foster for discussions, J. Sanchez-Yamagishi and P. Jarillo-Herrero for fabrication advice regarding the PVA, and I. Meric, Z. Kagan, A. Tsoi, N. Baklitskaya, and I. Mendonca for help with the device preparation. A portion of this work was performed at the National High Magnetic Field Laboratory, which is supported by National Science Foundation Cooperative Agreement No. DMR-0654118, the State of Florida and the U.S. Department of Energy.  This work is supported by DARPA CERA, AFOSR MURI, ONR MURI, FCRP through C2S2 and FENA, NSEC (No. CHE-0117752) and NYSTAR.

\bigskip
\section{Author Contributions}
CRD and AFY performed all experiments including sample fabrication and measurement, and wrote the paper. PC-Z contributed to sample measurement. LW and HR contributed to sample fabrication. KW and TT synthesized the h-BN samples. JH, PK, and KLS advised on experiments.

\end{document}